# Wavelength-stepping algorithm for testing thickness, front and back surfaces of optical plates with high signal-to-noise ratio

**MANUEL SERVIN,\* GONZALO PAEZ, MOISES PADILLA, AND GUILLERMO GARNICA**

*Centro de Investigaciones en Optica A. C., Loma del Bosque 115, Col. Lomas del Campestre, 37150 Leon Guanajuato, Mexico.*
*\*mservin@cio.mx*

We propose a least-squares phase-stepping algorithm (LS-PSA) consisting of only 14 steps for high-quality optical plate testing. Optical plate testing produces an infinite number of simultaneous fringe patterns due to multiple reflections. However, because of the small reflection of common optical materials, only a few simultaneous fringes have amplitudes above the measuring noise. From these fringes, only the variations of the plate's surfaces and thickness are of interest. To measure these plates, one must use wavelength-stepping, which corresponds to phase-stepping in standard digital interferometry. The designed PSA must phase-demodulate a single fringe sequence and filter out the remaining temporal fringes. In the available literature, researchers have adapted PSAs to the dimensions of particular plates. As a consequence, there are as many PSAs published as different testing plate conditions. Moreover, these PSAs are designed with too many phase-steps to provide detuning robustness well above the required level. Instead, we mathematically prove that a single 14-step LS-PSA can adapt to several testing setups. As is well known, this 14-step LS-PSA has a maximum signal-to-noise ratio (SNR) and the highest harmonics rejection among any other 14-step PSA. Due to optical dispersion and experimental length measuring errors, the fringes may have a slight phase detuning. Using propagation error theory, we demonstrate that measuring distances with around 1% uncertainty produces a small and acceptable detuning error for the proposed 14-step LS-PSA.

## 1. INTRODUCTION

Optical plates are widely used for advanced microchip production and optical instruments for measuring, observation and testing. Therefore, gauging the quality of the surfaces and thickness variation of optical plates with nanometer resolution is paramount. Optical engineers have been using phase-stepping algorithms since 1974 [1,2]. Phase-stepping use a piezoelectric to phase-step the reference beam of an interferometer [1,2]. However, for simultaneously testing the surface and thickness variations of an optical plate one needs a wave-tuning (wavelength stepping) interferometer [3-5]. On the other hand, it is well known that parallel surfaces produce infinitely many reflections between its surfaces [6,7]. The best known optical instrument that uses multiple reflections is the Fabry-Perot spectroscope [6,7]. Usually the Fabry-Perot etalon is used as narrow optical transmission filter [6,7]. Multiple reflection interference may also be used in the reflection direction for measuring optical plates [8-31]. The reflecting complex temporal fringe pattern contains phases corresponding to the front, back and thickness variations. Using wavelength tuning as phase stepping, we measure each phase while rejecting all others as harmonics [8-31]. If the wave-stepping algorithm (a PSA) is not adequate, the demodulated surface or thickness variations would contain crosstaking phase-ripples from simultaneous spurious reflecting fringes [8-31].

For interferometric testing of a single mirror-surface, the PSA is independent of the experimental set-up [1,2]. For example, a 4-step PSA works equally well for testing an optical mirror, as well as for fringe projection profilometry (FPP) [1,2]. Thus the PSA do not depend explicitly on the experimental conditions [1,2]. In contrast for simultaneously measuring the surfaces and thickness of plates, the designed PSA depends on the experimental set-up [8-32]. In other words, researchers have published as many PSAs as optical-plate testing experiments [8-32]. Early plate-testing experiments, only phase-demodulate the front flat-phase, filtering-out the other simultaneous fringes [8]. In this work, we derive two PSAs that work well for many experimental set-ups for testing the two surfaces and thickness variations of optical plates. The first phase demodulation requires only a 14-step least-squares PSA (LS-PSA) [1,2]. The second PSA is a 27-step PSA which is more robust to phase-detuning [2,8,11,34,35].

## 2. MATHEMATICAL MODEL FOR WAVE-STEPPING OPTICAL-PLATES TESTING

The mathematical theory for analyzing multiple reflections between parallel optical surfaces has been known for more than 100 years [3-7]. However, until 2000, multiple reflection theory was not used for optical-plate testing. In 2000 de Groot published a seminal paper for testing optical plates using a laser Fizeau interferometer [8]. The basics of the experimental set-up for testing optical plates by multiple reflections is shown in Fig. 1

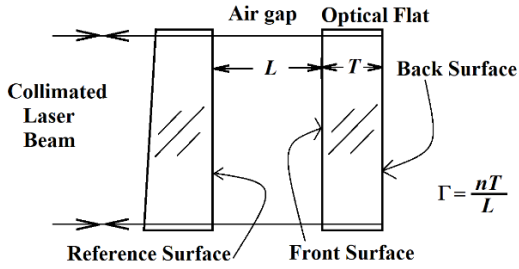

Fig. 1. Optical plate testing set-up showing the reference surface, the air-gap distance $L$, the plate thickness $T$ and refractive index $n$. Peter de Groot defined the useful optical length ratio $\Gamma=nT/L$ [8]. The value $\Gamma$ is chosen according to the designed PSA, and the air gap in the experiment is adjusted to $L=nT/\Gamma$ [8].

Figure 1 shows the basics for testing an optical-plate using a laser Fizeau interferometer. Following de Groot [8], each temporal interferogram $I(m)=I(x,y;m)$ can be rewritten as

$$I(x,y;m) = |u_n(x,y;m)|^2;$$
$$u(x,y;m) = \frac{r_0 + r'(m)\exp[i(\theta+v_1 m)]}{1+r'(m)r_0 \exp[i(\theta+v_1 m)]}; \quad (1)$$
$$r'(x,y;m) = \frac{r_1 + r_2 \exp[i(\varphi+\Gamma v_1 m)]}{1+r_1 r_2 \exp[i(\varphi+\Gamma v_1 m)]}.$$

Being the front surface phase delay $\theta(x,y)$ and the back phase $\varphi(x,y)$, both respect to the reference surface. Some $(x,y)$ coordinates were omitted for clarity. Constants $\{r_0, r_1, r_2\}$ are the complex reflection coefficients of the reference, front and back surfaces respectively. Usually, all three magnitudes are assumed equal and expressed in terms of the reflectance $R$ as $r_0 = -\sqrt{R}$, $r_1 = \sqrt{R}$, $r_2 = -\sqrt{R}$. The experimental parameters are the fundamental wave-stepping frequency $v_1$ and the optical-length ratio $\Gamma$; they are given by,

$$v_1 = 4\pi L \frac{\Delta\lambda}{\lambda^2}; \quad \text{and} \quad \Gamma = \frac{nT}{L}. \quad (2)$$

Where $\lambda$ is the central wavelength, and $\Delta\lambda$ the wavelength-step per temporal interferogram [8]. Equation (1) represents an infinite number of reflections interfering. However due to the small reflection coefficient (usually $R<0.1$, $n<2.0$), the number of visible harmonics is relatively low. Figure 2 shows the kind of reflecting fringe pattern at a pixel $(x,y)$ for a reflectance $R=0.1$, and $\Gamma=3.0$.

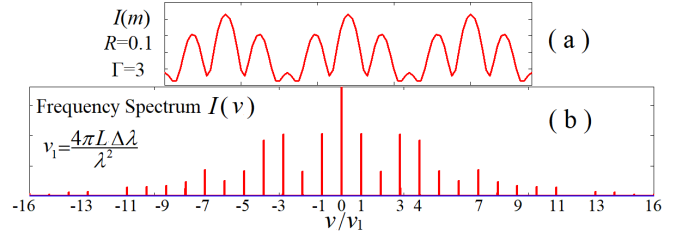

Fig. 2. A large temporal sequence of $I(m)$ (Eq. (1)), and its Fourier spectrum normalized to $v_1$. This spectrum has about 29 Dirac deltas above the measuring noise. The front, back and thickness phases are located at frequencies {1,3,4} respectively [8]. This shows the complexity of this interferometric problem as we need to extract the front, back and thickness phases from many crosstalking harmonics.

Figure 2(a) shows a temporal sequence of a multiple reflections interferogram (Eq. (1)) for an optical material with reflectance $R=0.1$, and optical-thickness to air-gap of $\Gamma=nT/L=3.0$. Equation (1) gives an infinite number of harmonic frequencies. However, as seen in Fig. 2(b), the number of harmonics above the noise level (typically -30 dB RMS for 8-bit cameras) are about 28 at both sides of the spectrum. From these harmonics, we must keep a single frequency and filter-out the remaining 27 harmonic frequencies. This is similar to FM radio broadcasting, where one is interested in tuning a single radio station and filtering-out all other FM stations.

## 3. SPECTRUM OF THE TEMPORAL FRINGES FOR MULTIPLE-REFLECTION INTERFERENCE

In order to analyze the harmonic content of multiple reflections interferograms, it is useful to express Eq. (1) in its cosine expansion [8,9]:

$$I(x,y;m) = a + \sum_{k=1}^{\infty} b_k \cos[\alpha_k + mv_k],$$
$$= a + \sum_{\substack{k=-\infty \\ k \neq 0}}^{\infty} \frac{b_k}{2} e^{i(\alpha_k + mv_k)}; \quad Z_k = \frac{b_k}{2} e^{i\alpha_k}. \quad (3)$$

One of the advantages offered by this series representation is that shows which phase variation $\alpha_k$ is located at the $k$-harmonic, with amplitude $b_k$ and at frequency $v_k$; finally, $a$ accounts for the illumination background of the fringes. The first nine strongest harmonics of Eq. (3) (also of Eq. (1)) are shown in Table 1 [8,9]. Considering the material reflectance around $R=0.04$ ($n=1.5$), only these 9 harmonics are above the noise level of the interferogram images [8,9].

**Table 1. Strongest 9 reflected signals and their temporal frequencies. The front phase is $\theta(x,y)$, the back phase is $\varphi(x,y)$, and thickness phase is $\theta(x,y)+\varphi(x,y)$. [8,9].**

| Harmonic $k$ | Frequency $v_k/v_1$ | Amplitude $b_k/\|b_1\|$ | Phase $\alpha_k$ |
|---|---|---|---|
| 1 | 1 | -1 | $\theta$ |
| 2 | $\Gamma$ | -1 | $\varphi$ |
| 3 | $\Gamma+1$ | 1 | $\theta+\varphi$ |
| 4 | 2 | -R | $2\theta$ |
| 5 | $\Gamma-1$ | -R | $\varphi-\theta$ |
| 6 | $\Gamma+2$ | 2R | $2\theta+\varphi$ |
| 7 | $2\Gamma$ | -R | $2\varphi$ |
| 8 | $2\Gamma+1$ | 2R | $\theta+2\varphi$ |
| 9 | $2\Gamma+2$ | -R | $2\theta+2\varphi$ |

As Table 1 shows, the three searched phases are the front plate surface $\theta(x,y)$, the back surface phase $\varphi(x,y)$, and the thickness variation $\theta(x,y)+\varphi(x,y)$. Therefore, knowing any two of these phases one may know the third one.

Figure 3 shows the fast Fourier transform (FFT) spectra of a long temporal sequence of Eq. (1), with $R=0.04$ and $\Gamma=(nT/L)=\{1/10, 1/3, 3, 10\}$. It is particularly important to note that the position of the harmonic changes for each $\Gamma$ value (optical-length ratio). For this reason, it is not obvious to choose a single PSA for each change of the experimental conditions stated by $\Gamma$. That is why, researchers have published as many PSAs as optical-plate testing experiments.

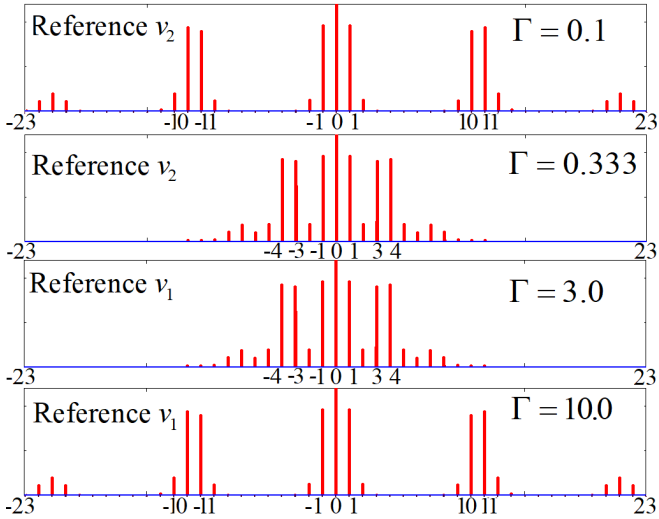

Fig. 3. Fast Fourier transform (FFT) of large fringe sequences $I(m)$ for $\Gamma=\{1/10, 1/3, 3, 10\}$ and $R=0.04$. Note that $\Gamma=\{1/10, 10\}$, and $\Gamma=\{1/3, 3\}$ have the same FFT, but their fundamental frequency are different; $v_2$ and $v_1$ respectively. The first harmonic for $\Gamma=1/10$ contains the back surface phase, while its 10th harmonic contains the front phase.

Figure 3 shows a couple of $\Gamma$ values and their reciprocals. We have found a useful fact that the reciprocal of the $\Gamma$ parameter has the same spectra whenever the reference frequency is changed. Therefore we may use the same PSA for filtering the plate surfaces for $\Gamma=\{1/10, 10\}$, and $\Gamma=\{1/3, 3\}$. The interesting fact of $\Gamma=1/10$, is that one has an air-gap distance 15 times longer ($L=nT/\Gamma$, $n=1.5$) than the physical thickness of the plate. This is an experimental advantage for testing thin optical plates

## 4. MODULUS OF THE ANALYTIC SIGNAL FOR HARMONIC REJECTION GAUGING

Here we mathematically prove that the modulus of the filtered analytic signal may be used to gauge the robustness of harmonic rejection. As Fig. 1 and Eq. (1) show, the phases $\theta(x,y)$, $\varphi(x,y)$, and $\theta(x,y)+\varphi(x,y)$ are the front, back and thickness variations. The testing experiment fixes the parameters $\Gamma$, and $R$. Then one takes a temporal sequence of $N$ fringes $I(x,y;m)$ to demodulate, let say $\theta(x,y)$ as,

$$Z = \sum_{m=0}^{N-1} c_m I(x,y;m); \quad c_m \in \mathbb{C}. \quad (4)$$

Ideally, $Z$ may only contain the front-surface phase, $|Z_1|\exp[i\theta(x,y)]$. However, if the PSA is not the appropriate one, or the fringes are detuned, $Z$ would contain corrupting crosstalking harmonics [32]. Therefore, one may obtain, for example, the following analytic signal,

$$Z = |Z_1|e^{i\theta} + |Z_{-1}|e^{-i\theta} + |Z_{-6}|e^{-i(2\theta+\varphi)} + |Z_9|e^{i(2\theta+2\varphi)}. \quad (5)$$

In this example, the desired signal is $|Z_1|e^{i\varphi}$, but it is corrupted by crosstalking harmonics at frequencies {-1,-6,+9}. In the optical shop testing, how can we know that spurious harmonics are present? Experimentally we cannot separate the spurious harmonics from the desired signal $|Z_1|e^{i\theta(x,y)}$ as we do in computer simulations.

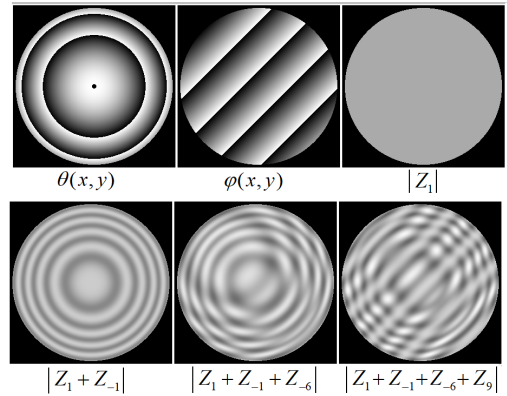

Fig. 4. The first row show the front phase $\theta(x,y)$, the back phase $\varphi(x,y)$, and the correctly demodulated modulus $|Z_1|$, which shows no fringe structure from crosstalking harmonics. The second row shows standard double-fringe detuning structure $|Z_1+Z_{-1}|$. The images $|Z_1+Z_{-1}+Z_{-6}|$, and $|Z_1+Z_{-1}+Z_{-6}+Z_9|$ show increasing fringe structures [32].

A way out of this conundrum is to picture (see Fig. 4) the modulus of $|Z|$ in Eq. (5). If $|Z|$ has no fringe structure, then we are absolutely sure that no corrupting harmonics are present [32]. For the sake of clarity let us further exemplify our PSA outcome in Eq. (5) as.

$$Z = Z_{+1} + Z_{-1} + Z_{-6} + Z_{+9} ;$$
$$Z = e^{i\theta} + 0.1e^{-i\theta} + 0.07e^{-i(2\theta+\varphi)} + 0.05e^{i(2\theta+2\varphi)}. \quad (6)$$

In Eq. (6) the desired signal is $Z_{+1}$, but we may end up with spurious signals, $Z_{-1} + Z_{-6} + Z_{+9}$. Figure 4 shows how $|Z|$ (Eq. (6)) would look like, as we increment the number of crosstalking signals.

## 5. LEAST-SQUARES PSA FOR OPTICAL-PLATE TESTING

The first $N$-step PSA was proposed by Bruning et al. [1]; here we refer to this PSA as the least-squares PSA (LS-PSA). The LS-PSA has the highest signal-to-noise ratio (SNR=$N$), and the highest harmonic rejection capacity for a given number $N$ of phase steps [2]. An important problem for testing optical flat plates is the high number of cross-talking harmonics of the fringes, as can be seen in Eq. (6) and showed at Fig. 2 and Fig. 4. Here we show that a LS-PSA is well suited to deal with the large amount of harmonics caused by the multiple reflections interference. Furthermore, we have found the minimum number of phase steps needed to separate the front, back and thickness phase signals. This minimum-steps LS-PSA may be frequency-shifted and tuned at the three different frequencies where these signals are located.

We can take a sequence of $N$ fringes $I(x,y;m)$ to demodulate the front surface phase $\theta(x,y)$ as,

$$|Z(x,y)|e^{i\theta(x,y)} = \sum_{m=0}^{N-1} e^{-imv_1} I(x,y;m). \quad (7)$$

Being $v_1$ the phase-step between consecutive interferograms. According to Eq. (2) the numerical phase-step $v_1$ must coincide with the experimental one; that is,

$$v_1 = 4\pi L \frac{\Delta\lambda}{\lambda^2} = \frac{2\pi}{N} \quad \Rightarrow \quad \Delta\lambda = \frac{\lambda^2}{2LN}. \quad (8)$$

This equation gives the wavelength-stepping size $\Delta\lambda$ of the plate-testing experiment.

As Fig. 4 shows, the highest frequency harmonics is 22 for the case $\Gamma=\{0.1,10\}$. To reject 22 harmonics using a LS-PSA, it would require at least 24 phase steps [2]. However, because these spectra are sparse (there are many harmonic frequencies with zero energy), we propose to use 14-steps PSA as Fig. 5 shows. Here the 14-step PSA's frequency transfer function (FTF) is represented by $H_{14}(v)$ [2]. Even further, with the same 14-step phase-shifted interferogram data, we can found the front, back and thickness phases by translating the FTF (frequency shifting) the same 14-step LS-PSA. This is shown in Fig. 5 for $\Gamma=0.1$.

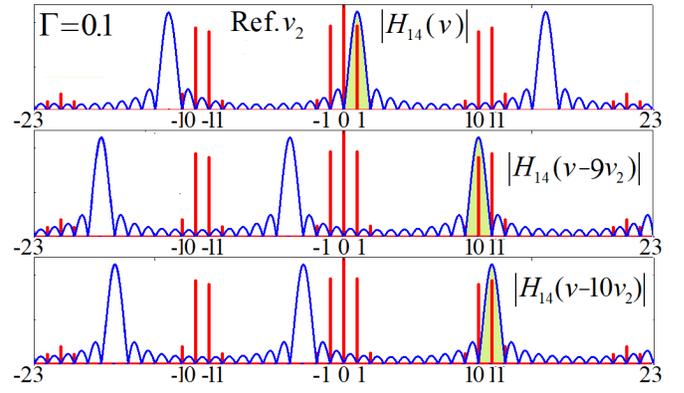

Fig. 5. The proposed 14-step LS-PSA with a phase-step of $2\pi/14$ and frequency displaced. The first FTF $H_{14}(v)$ demodulates the front-surface harmonic. The same and displaced $H_{14}(v-9v_2)$ demodulates the back-surface. Finally, $H_{14}(v-10v_2)$ demodulates the thickness variation. Note that the sparsity of spectra allows us to use just 14 steps ($R=0.04$).

Figure 6 shows that the same 14-step LS-PSA is capable of phase demodulating the front surface variation for all $\Gamma=\{0.1, 0.333, 3.0, 10\}$.

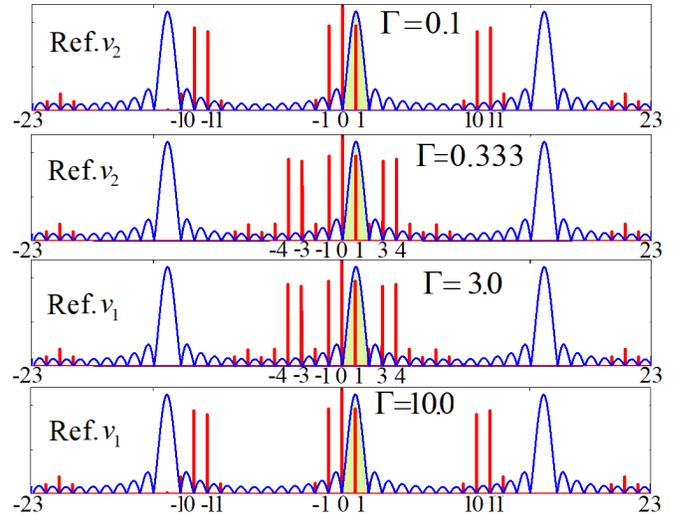

Fig. 6. Same 14-step LS-PSA (with different fundamental frequency) is used for demodulating the front surface phase for experimental set-ups with $\Gamma=\{0.1, 1/3, 3.0, 10.0\}$. The filtered signal is the fundamental one ($R=0.04$).

Next Fig. 7 shows the proposed 14-step LS-PSA's FTF displaced to frequencies $\{v_1, 3v_1, 4v_1\}$ for $\Gamma=3.0$. The analytic signal $Z_1$ corresponding to the front surface-phase is located at $v_1$. The analytic signal $Z_2$ corresponding to the back surface-phase is located at $3v_1$. And finally, the analytic signal $Z_3$ corresponding to the thickness-phase is located at $4v_1$.

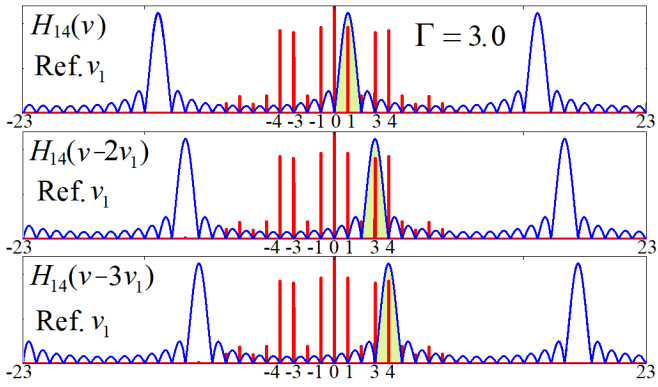

Fig. 7. Same 14-step PSA with a phase-step of $2\pi/14$, but frequency displaced. The FTF $H_{14}(v)$ demodulates the front surface of the plate. This FTF but centered at $3v_1$ and $4v_1$ demodulates the phase of the back and thickness variations of the optical-plate ($R$=0.04).

Figure 8 shows full harmonic rejection (no fringe structure for $|Z_1|$, $|Z_2|$ and $|Z_3|$) of well-tuned 14-steps LS-PSA's phases corresponding to $\Gamma=\{1/3, 3\}$.

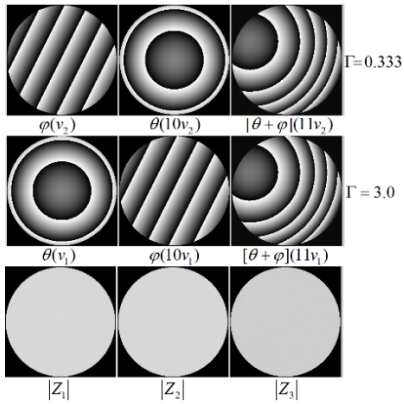

Fig. 8. The upper row shows (well-tuned) demodulated phases for the back, front, and thickness phases for $\Gamma=1/3$. The middle row shows the front, back, and thickness phase for $\Gamma=3.0$. Note that the front and back phases are interchanged. The bottom row shows $|Z_1|$, $|Z_2|$, $|Z_3|$ with no fringe error demodulation structure due to crosstalking fringes.

Figure 9 shows the fringes structure crosstalking for a 1% detuning of the $\Gamma$ parameter using the 14-step LS-PSA.

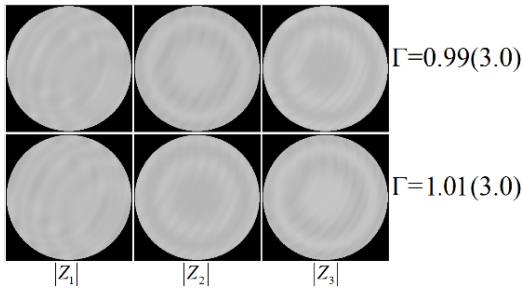

Fig. 9. Fourteen-steps LS-PSA tolerates up to 1% of measuring error deviation for $\Gamma=nT/L$ in Eq. (1) to obtain reliable phase demodulation.

## 6. PHASE DETUNING DUE TO OPTICAL DISPERSION AND MEASURING ERROR OF Γ

In this section we mathematically prove the detuning error sensitivity due to the optical material light-dispersion, and the measuring distance errors for $T$ and $L$.

### 6.1 Detuning due to light dispersion in wavelength stepping

Let us assume the same experimental conditions as de Groot [8]. Using a laser wavelength of $\lambda$=680nm, and the thickness of the optical plate is $T$=10cm. The Sellmeier equation for BK7 optical glass is given by,

$$n_{BK7}(\lambda) = \sqrt{1 + \frac{a\lambda^2}{\lambda^2 - b} + \frac{c\lambda^2}{\lambda^2 - d} + \frac{e\lambda^2}{\lambda^2 - f}}. \quad (9)$$

Being $a$=1.03961212, $b$=0.00600069867, $c$=0.231792344, $d$=0.0200179144, $e$=1.01046945, and $f$=103.560653. From Eq. (2), the air-gap distance is $L = n_{BK7}(\lambda)T/\Gamma$ = $(0.01)n_{BK7}(0.680)/3.0$ =5.04cm. Then, the wavelength-stepping for a 14-step LS-PSA must be (see Eq. (8))

$$\Delta\lambda = \frac{\lambda^2}{2LN} = \frac{(680\times 10^{-9})^2}{2(0.0504)(14)}10^9 \text{ nm} = 3.277\times 10^{-4} \text{ nm}. \quad (10)$$

The total variation of the index of refraction of the BK7 due to 14 and 27-steps $\Delta\lambda$ is computed as.

$$n_{BK7}(\lambda)=1.51361483;$$
$$n_{BK7}(\lambda+13\Delta\lambda)=1.51361471; \quad (11)$$
$$n_{BK7}(\lambda+26\Delta\lambda)=1.51361459.$$

The refractive index of BK7 glass $n_{BK7}(\lambda+m\Delta\lambda)$ does not change up to the 7th significant figure even for 26 wavelength steps. The $\Gamma$ parameter in Eq. (1) varies linearly with the refractive index as $\Gamma = n_{BK7}(\lambda+26\Delta\lambda)T/L = 3.00000$, as a consequence, the $\Gamma$ variation produces no significant detuning error for a 14-step LS-PSA.

In this work we are considering phase stepping, not phase-shifting integration bucket [8]. Thus, the $N$ wavelength-stepped interferograms with increments $\Delta\lambda$ translate into a phase-step of $2\pi/N$ (Eq. (8)); this is modeled as

$$\sum_{m=0}^{N-1} I(x,y;\lambda+m\Delta\lambda)\delta(t+m\Delta t) \quad (12)$$

Being $I(x,y;\lambda+m\Delta\lambda)$ the interferogram image at constant wavelength $(\lambda+m\Delta\lambda)$, at sampling time $\delta(t+m\Delta t)$ (see Fig. 10). If a laser undergoes a wavelength-shift from i.e. $\lambda$ to $(\lambda+\Delta\lambda)$ between times $\delta(t)$ and $\delta(t+\Delta t)$, the instantaneous wavelength must change $|d\lambda/dt|>0$ within the open interval $(t, t+\Delta t)$. However, here we are assuming that the wavelength $(\lambda+m\Delta\lambda)$ is held constant at instants $\delta(t+m\Delta t)$. Therefore, as Fig. 10 shows, $(d\lambda/dt)$ plays no role in the present work, as it was the case for Ref. [10].

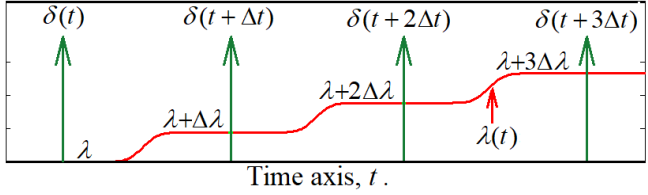

Fig. 10. A possible wavelength variation with time. The wavelength variation is the red trace, and the fringe instantaneous samplings are represented by green Dirac deltas.

### 6.2 Detuning due to measuring physical distances uncertainty

On the other hand, $\Gamma = n_{BK7}T/L$ may vary more significantly than the index $n_{BK7}$ (see Eq. (9)). When one chose $\Gamma=3.0$, and measuring a plate with thickness of $T=100$mm, one calculates that the air-gap is $L= n_{BK7}T/\Gamma =50$mm. If $T$ and $L$ are measured with a Mitutoyo caliper (Model 573-621-20) having an uncertainty of $\sigma=0.02$mm. The error-propagation formula for $\Gamma = n_{BK7}T/L$ is [33],

$$\sigma_\Gamma = \sqrt{\left(\frac{\partial \Gamma}{\partial T}\sigma\right)^2 + \left(\frac{\partial \Gamma}{\partial L}\sigma\right)^2} = \Gamma\sqrt{\left(\frac{\sigma}{T}\right)^2 + \left(\frac{\sigma}{L}\right)^2};$$

$$\frac{\sigma_\Gamma}{\Gamma} = \sqrt{\left(\frac{0.02}{100}\right)^2 + \left(\frac{0.02}{50}\right)^2} = 0.00045 = 0.045\%. \quad (13)$$

This is the percentage of error for measuring $\Gamma$ (with $n_{BK7}=1.5$). In the previous section we saw that the 14-step LS-PSA may tolerate up to 1% of error for $\Gamma$. By propagating the measuring errors using Eq. (10), the fractional error $\sigma_\Gamma/\Gamma = 0.00045$, or 0.045%. This is well below the 1% upper-bound tolerance seen in Fig. 9.

On the other hand, now assume a thickness of $T=10$mm (10 times thinner than previous case). If we use $\Gamma=3$ as before, the air gap would be $L=(n_{BK7}T/\Gamma)=5.045$ mm. Working with such narrow air gap may be difficult. Instead, we choose $\Gamma=1/3$, thus $L=(n_{BK7}T/\Gamma)=45.41$mm; this wider air gap is easier to implement in the experiment. Again $T$ and $L$ are measured with a Mitutoyo caliper with an uncertainty of $\sigma=0.02$mm. The error-propagation formula now gives,

$$\frac{\sigma_\Gamma}{\Gamma} = \sqrt{\left(\frac{0.02}{10}\right)^2 + \left(\frac{0.02}{45}\right)^2} = 0.002 = 0.2\%. \quad (14)$$

For this wider air gap, the fractional error increases to $\sigma_\Gamma/\Gamma = 0.002$, or 0.2%. This is however five times lower than 1%, which is the upper bound for a 14-step LS-PSA to obtain very low crosstalking detuning error (see Fig. 9). In other words, the measuring uncertainty could be as high as 0.1 mm and the 14-step LS-PSA would still have small detuning error.

## 7. PHASE DEMODULATION WITH A 27-STEPS DETUNING ROBUST PSA

The 14-step LS-PSA has the highest possible SNR=14, and the highest harmonic rejection [2]. However, if the relative error $\sigma_\Gamma/\Gamma$ is larger than 1%, then phase detuning error would appear in the 14-step LS-PSA. Therefore, here we propose to increase the detuning robustness of the LS-PSA designing a detuning-robust 27-step PSA as follows.

The frequency-transfer-function (FTF) of a 14 step LS-PSA (the only PSA we have used so far) is,

$$H_{14}(v) = \sum_{m=0}^{13} e^{-im\frac{2\pi}{14}} e^{imv} = \prod_{m=0}^{12}\left[1 - e^{i\left(v - m\frac{2\pi}{14}\right)}\right]. \quad (15)$$

Being $e^{imv} = F\{\delta(t-m)\}$ and $F[.]$ is the Fourier transform operator. If we square this FTF, one obtains a 27-step PSA with robust-detuning, second order zeroes as,

$$H_{27}(v) = \left\{\sum_{m=0}^{13} e^{-im\frac{2\pi}{14}} e^{imv}\right\}^2 = \prod_{m=0}^{12}\left[1 - e^{i\left(v - m\frac{2\pi}{14}\right)}\right]^2. \quad (16)$$

Convolving the square-window 14-step $H_{14}(v)$ with itself, one obtains a 27-step PSA. The envelope window of this 27-step PSA $H_{27}(v)$ is a triangle. See Fig. 11 and Fig. 12.

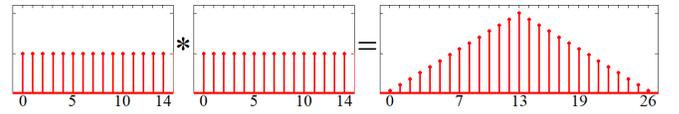

Fig. 11. Self-convolution of a LS-PSA square-window. The 14-step LS-PSA is self-convolved to obtain a detuning robust 27-steps PSA with triangle-weighted window.

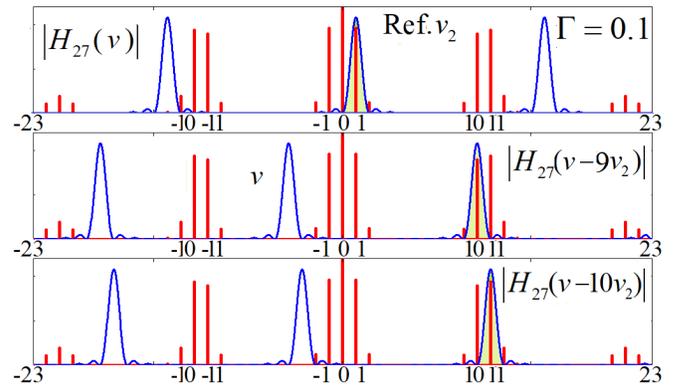

Fig. 12. Detuning-robust 27-step PSA. These three FTFs are all equal, but located at frequencies $\{v_2, 10v_2, 11v_2\}$. The fundamental frequency is $v_2$ ($R$=0.04).

Figure 12 shows a detuning robust 27-step PSA for $\Gamma=0.1$ obtained by squaring the FTF of the 14-step LS-PSA. For other values of Γ, as shown in previous sections, we use the same 27-step PSA but displaced in frequency as many times as required to filter the desired harmonic.

Figure 13 shows how the fringe spectra are expanded or compressed (detuned) when Γ varies up to 5%. Even with this frequency displacement (due to detuning), the crosstalking harmonics are fully rejected.

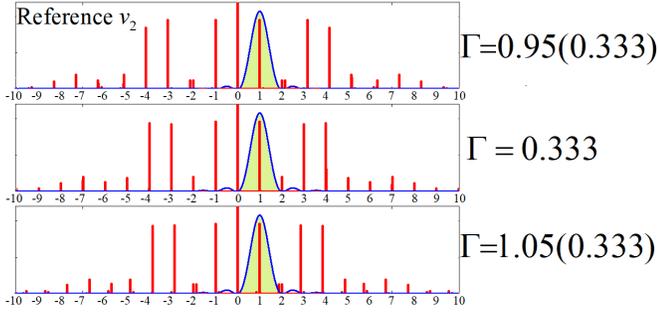

Fig. 13. The fringe spectra are compressed or expanded as Γ=$nT/L$ varies. The detuning robustness of this 27-step PSA makes that most harmonic energy do not pass through the quadrature filtering ($R$=0.04).

Figure 14 shows the modulus fringe-error structure of the 27-step PSA for a 5% in variation for Γ. The upper row in Fig. 14 shows the back, front and thickness phase variations for detuned Γ=0.95(0.3333) fringes. The lower row in Fig. 14 shows the back, front and thickness phase variations for detuned Γ=1.05(0.3333) fringes. In the upper row (Fig. 14) shows, the back and front phases as the best estimates, showing no fringe structure. In contrast, for the lower row, the best phase estimations are the back and thickness phases; they both show no fringe structure in their modulus.

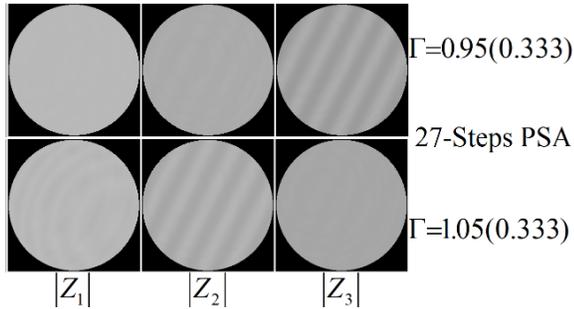

Fig. 14. Upper row (Γ=0.95×(1/3)), from left to right shows, the front, back and thickness analytic-signal modulus; the back and front phases are the most reliable ($|Z_1|,|Z_2|$). On the other hand, the lower row (Γ=1.05×(1/3)), shows the front and thickness phases ($|Z_1|,|Z_3|$) as the best estimation ($R$=0.04).

Thus we can say that if our optical length ratio Γ=$nT/L$ is well measured (with less than 1% uncertainty), then a 14-step LS-PSA gives good estimations of the optical plate. If this is not possible, then one may use the 27-steps PSA recommended in this section.

Finally, the SNR for the 27-steps PSA is given by,

$$\text{SNR}_{27} = \frac{|H_{27}(v_2)|^2}{\frac{1}{2\pi}\int_{-\pi}^{\pi}|H_{27}(v)|^2 dv} = 20.222. \quad (17)$$

This figure-of-merit is higher than the SNR reported for other detuned-robust PSAs [10-31]. The $\text{SNR}_{27}$ is independent of the frequency displacement of the PSA to demodulate the three signals of interest. Table 2 extends the comparison published by Jeon et al. [25].

**Table 2. SNR for some PSAs. We define the SNR-efficiency as SNR/steps-number=SNR/$N$. The proposed LS-PSA is 100% efficient (SNR/$N$=1.0).**

| PSA $N$-steps | SNR | Efficiency (SNR/$N$) | Year |
|---|---|---|---|
| Proposed 14-steps | 14.000 | 1.000 | 2023 |
| Proposed 27-steps | 20.222 | 0.749 | 2023 |
| Estrada 9-steps [35] | 5.963 | 0.663 | 2009 |
| Jeon 21-steps [25] | 13.864 | 0.660 | 2022 |
| Jeon 11-steps [24] | 7.111 | 0.646 | 2021 |
| Hibino 11-steps [34] | 5.102 | 0.464 | 1995 |
| Kim 13-steps [20] | 5.044 | 0.388 | 2020 |

Table 2 shows the SNR of some PSAs with a new figure of merit, the SNR-efficiency=(SNR/$N$). The $N$-step algorithm with the highest SNR is the LS-PSA which is 100% efficient (SNR/$N$=1.0).

## 8. CONCLUSION

We mathematically proved that LS-PSA allows simultaneous measuring of front, back and thickness variation of optical plates using multiple-reflections interferometry. Particularly, we proposed a 14-step LS-PSA that adapts to several plate-testing geometries. In previous works [10-31], researchers used a much higher phase-step number to increase the detuning robustness of their PSAs. But here we mathematically prove that such high detuning robustness is often unnecessary. We next highlight other contributions made in this work:

- We have shown that the spectra corresponding to Γ and 1/Γ (Eq. 1) have the same harmonics distribution (Fig. 3).
- We have shown that the same 14-step LS-PSA may be used with common Γ values {1/3, 3} as well as faraway Γ values as {1/10, 10}. These faraway Γ values become handy for testing thinner or thicker plates.
- We use the modulus of the demodulated analytic signal to gauge the rejection of simultaneous crosstalking harmonics.
- We analyzed the uncertainty propagation towards Γ that produces fringe detuning.
- We show that a 14-step LS-PSA gives good phase estimations of the plate when Γ is measured with less than 1% uncertainty.
- We designed a detuning-robust 27-step PSA which tolerates up-to 5% for Γ uncertainty.
- Both proposed PSAs have a high SNR-efficiency as Table 2 shows.

- We have shown that Γ values of {1/10, 10} translate into a much wider and sparse spectra. Due to this sparsity it is possible to use few-step PSA merely displacing it to the harmonics of interest.

Finally, we wish to remark that if we have a systematic bias in Γ, due to optical dispersion as in Ref. [10], then we only need to readjust the air gap $L=nT/\Gamma$ to compensate this new value of Γ.

**Acknowledgments.** The authors thank the Mexican Council for Science and Technology (CONACYT) for general support.

**Disclosures.** The authors declare no conflict of interest.

**Data Availability Statement.** No experimental data were generated in the presented research.